\documentclass[aps,prd,preprint,nofootinbib]{revtex4-1}
\usepackage{amsfonts}
\usepackage{mathrsfs}
\usepackage{graphicx}
\usepackage{amsmath}
\usepackage{amssymb}
\usepackage{graphicx}
\usepackage{slashed}
\usepackage{color}
\usepackage{multirow}
\usepackage{ulem}
\usepackage{adjustbox}

\usepackage{stmaryrd}
\newcommand{\bqa}{\begin{eqnarray}}
	\newcommand{\eqa}{\end{eqnarray}}
\newcommand{\beq}{\begin{equation}}
	\newcommand{\eeq}{\end{equation}}
\allowdisplaybreaks[1]

\graphicspath{{Fig/}{dia/}} \DeclareGraphicsExtensions{.eps}
\begin{document}
	
	\title{Heavy quark symmetry of $\Lambda_b^0$ decays in quark models}
	
	\author {Jiabao Zhang\footnote{zhangjiabao21@mails.ucas.ac.cn}, Xiang-Nan Jin\footnote{jinxiangnan21@mails.ucas.ac.cn}, Chia-Wei Liu\footnote{chiaweiliu@ucas.ac.cn} and Chao-Qiang Geng\footnote{cqgeng@ucas.ac.cn}}
	\affiliation{School of Fundamental Physics and Mathematical Sciences, Hangzhou Institute for Advanced Study,~UCAS, Hangzhou 310024, China\\
    University of Chinese Academy of Sciences, 100190 Beijing, China
	}
	\date{\today}
	
\begin{abstract}
We study the heavy quark symmetry with the homogeneous bag model~(HBM) and light-front quark model~(LFQM) based on the decays of $\Lambda_b^0\to\Lambda_c^+\ell^-\overline{\nu}_\ell~(\ell=e,\mu,\tau)$.
In particular, we calculate various parameters in the heavy quark expansions, including the Isgur-Wise functions and their first order corrections. 
The parameters in the HBM are fitted from the mass spectra, while the ones in the LFQM are tightly constrained by the heavy quark symmetry, granting the predictive power of our results. 
We explicitly obtain that ${\cal B}(\Lambda_b^0\to\Lambda_c^+e^-\overline{\nu}_e)=(5.69\pm 0.58, 5.35\pm 0.50)$, ${\cal B}(\Lambda_b^0\to\Lambda_c^+\mu^-\overline{\nu}_\mu)=(5.67\pm 0.58 $, $5.33\pm 0.49)$, and $\Gamma(\Lambda_b^0\to\Lambda_c^+\tau^-\overline{\nu}_\tau)/\Gamma(\Lambda_b^0\to\Lambda_c^+\mu^-\overline{\nu}_\mu)=(0.3243\pm 0.0126$, $ 0.3506\pm 0.0046)$ for the numerical values of (HBM, LFQM). Our results of the branching fractions in both models agree well with the experimental data and lattice QCD calculations.
In addition, we find that the hard gluon corrections decrease the branching fractions around $10\%$.
\end{abstract}
	
	\maketitle
	
	\section{Introductions}
	
\begin{figure}[b]
		\centering
		\includegraphics[width=0.45\linewidth]{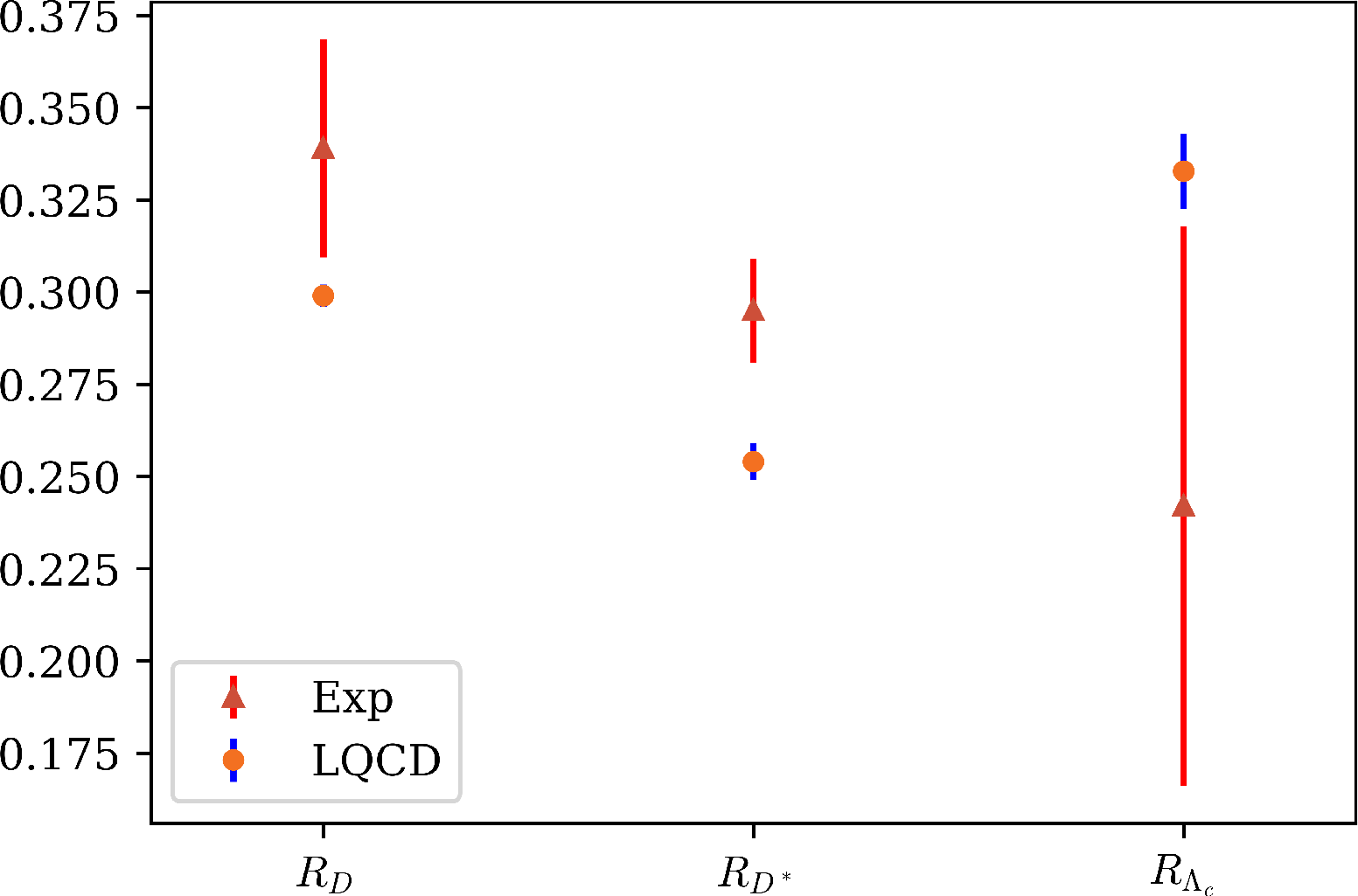}
		\caption{The ratios of $R_D,R_{D^*}$ and $R_{\Lambda_c}$ in the experiments and LQCD.}
		\label{moti}
	\end{figure}

	Testing the lepton universality via beauty quark decays has raised great interest in both theories and experiments~\cite{LHCb:2021trn,LHCb:2022wrs,Iguro:2022yzr,Choudhury:2022ktg,Belle:2021dgc,Sheng:2020drc,Das:2019cpt,Zhang:2019xdm,Groote:2021ayy,Azizi:2018axf,DiSalvo:2018ngq,Gutsche:2015mxa}.
	Recently, the  LHCb Collaboration has reported the ratio of
	$R_{\Lambda_c}= {\cal B}(\Lambda_b^0\to\Lambda_c^+\tau^-\overline{\nu}_\tau)/ {\cal B}(\Lambda_b^0\to\Lambda_c^+\mu^-\overline{\nu}_\mu)
	$, given by~\cite{LHCb:2022piu,LHCb:2017vhq}
	\begin{equation}
		R_{\Lambda_c}=0.242\pm0.026\pm0.040\pm0.059,
	\end{equation}
	where the first and second uncertainties are systematic and statistical in 
	${\cal B}(\Lambda_b^0\to\Lambda_c^+\tau^-\overline{\nu}_\tau)$, and the third one comes from ${\cal B}( \Lambda_b^0\to\Lambda_c^+\mu^-\overline{\nu}_\mu)$, respectively. 
	On the other hand, the lattice quantum chromodynamics~(LQCD) gives a slighly larger ratio~\cite{Detmold:2015aaa,Bernlochner:2018kxh}. The results along with the meson versions are summarized in Fig.~\ref{moti} with $R_{D^{(\ast)}}=\Gamma(B\to D^{(\ast)} \tau^- \overline{\nu}_\tau)/\Gamma(B\to D^{(\ast)} e^- \overline{\nu}_e) $~\cite{HFLAV:2022pwe,Martinelli:2022fgg}.
	Notice that the experimental values of $R_{D^{(\ast)}}$ are larger than the theoretical ones in contrast to $R_{\Lambda_c}$.
	
	On the theoretical side, the form factors of $\Lambda_b^0\to\Lambda_c^+$ provide ideal playgrounds for quark models. The main reason is that at the massless limit of $(u,d)$, there are only two  energy scales in $\Lambda_Q$ with $Q=(b,c)$, given by
	\begin{equation}
		\overline{\Lambda} = \lim_{m_Q \to \infty} \left( M_Q - m_Q \right) \,,~~~\varepsilon_Q = \overline{\Lambda}/2m_Q \ll 1 \,,
	\end{equation}
	where $M_Q$ and $m_Q$ are the heavy baryon and quark masses, respectively. As $\varepsilon_Q$ are  tiny, the physical quantities are Taylor expanded regarding $\varepsilon_Q$, namely the heavy quark expansion~(HQE).

	We take the Isgur-Wise function as an illustration, given as 
	\begin{equation}\label{xifun}
		\overline{\xi}(\omega)=\xi(\omega)+(\varepsilon_b+\varepsilon_c)\xi_{ke}(\omega)+{\cal O}(\varepsilon_c^2)\,,
	\end{equation}
	where $\xi(\omega)$ is the Isgur-Wise function, $\xi_{ke}(\omega)$ is the first order correction, and 
	$\omega=v_b \cdot v_c$ with $v_Q$ the four-velocities of $\Lambda_Q$. 
	As we will see in the next section, $\overline{\xi}(\omega)$ governs the recoil effects of the form factors.
	Although the experiments can  probe $\overline{\xi}(\omega)$ only, Eq.~\eqref{xifun} allows us to compare $\xi(\omega)$ and $\xi_{ke}(\omega)$ separately among the quark  models, without the dependence of the quark masses.
	
	In this work, we take the homogeneous bag model~(HBM) and LFQM to illustrate the heavy quark symmetry.
	These models are unimpressive but interesting in different aspects.
	On the one hand, the HBM is a relativistic quark model, in which $u$ and $d$ quarks can be safely taken as massless, and the parameters are fitted from the  mass spectra. It is reliable at the zero recoil point $(\omega =1 )$, but unequal time commutators are needed for a boosted state, causing several uncertainties as $\omega$ goes up. 
	On the other hand, the LFQM describes the bound state in a frame-independent way, in which only the relative motions between the constituents are required~\cite{Schlumpf:1992vq,Cheng:1996if,Cheng:2004cc,Ke:2007tg,Ke:2012wa,Wang:2017mqp,Geng:2022xpn,Zhao:2022vfr,Chen:2021ywv,Chang:2018zjq,Chua:2019yqh}.
	The Lorentz boosts in the front form are generated by the kinematical operators, which leave the $x^+=0$ plane invariant~\cite{Dirac:1949cp,Chiu:2017ycx}, so the  unequal time commutation relations are not needed.
	However, the LFQM suffers the uncertainties from the parameters input. In addition, the $Z$-graph contributions forbid us from computing the form factors in the timelike region~\cite{ODonnell:1995dio,Simula:1996pk,Cardarelli:1997sx,Demchuk:1998gv,Choi:1998nf,Choi:2012zzb,Choi:2011xm}. We will show  
	that these aforementioned difficulties are resolved by the heavy quark symmetry. 
	
	This paper is organized as follows.
	We present the formalism of the HBM and LFQM in Sec.~\ref{sec2}.
	The numerical results are given in Sec.~\ref{sec3}.   Section \ref{sec4} is the conclusion.
	
	\section{Formalism}\label{sec2}
	We briefly review some of the results of the HQE, where the details can be found in Ref.~\cite{Bernlochner:2018kxh}.
In the heavy quark system, there are two popular parametrizations for the form factors, given as 
	\begin{align}
		\begin{aligned}
			\langle\Lambda_c^+|\overline{c}\gamma^\mu(1-\gamma_5)b|\Lambda_b^0\rangle=\overline{u}_{c}&\left[(F^V_1(\omega){\gamma^{\mu}}+F^V_2(\omega)v_b^\mu+F^V_3(\omega)v_c^\mu)\right.\\
			&~-\left.(F^A_1(\omega){\gamma^{\mu}}+F^A_2(\omega)v_b^\mu+F^A_3(\omega)v_c^\mu)\gamma_5\right]u_{b}\\
			=\overline{u}_c&[(f_1(q^2)\gamma^\mu-f_2(q^2)\frac{i\sigma^{\mu\nu}q_\nu}{M_b}+f_3(q^2)\frac{q^\mu}{M_b})\\
			&~-(g_1(q^2)\gamma^\mu-g_2(q^2)\frac{i\sigma^{\mu\nu}q_\nu}{M_b}+g_3(q^2)\frac{q^\mu}{M_b})\gamma_5]u_b,\\
		\end{aligned}
	\end{align}
	where $q= p_b - p_c $, and $p_b $ and $p_ c$ are the four momenta of $\Lambda_b^0$ and $\Lambda_c^+$, respectively. 
From $p_Q = m_Q v_Q$,	it is straightforward to show that $q^2=M_b^2+M_c^2-2M_b\,M_c\,\omega$.
	Two sets of  parametrizations are related as 
	\begin{equation}\label{Ff}
		\begin{aligned}
			&f_1=F_1^V+\frac{M_+}{2M_b}F_2^V+\frac{M_+}{2M_c}F_3^V,~&f_{2,3}=\mp\frac{1}{2}F_2^V-\frac{M_b}{2M_c}F_3^V,\\
			&g_1=F_1^A-\frac{M_-}{2M_b}F_2^A-\frac{M_-}{2M_c}F_3^A,~&g_{2,3}=\mp\frac{1}{2}F_2^A-\frac{M_b}{2M_c}F_3^A,\\
		\end{aligned}
	\end{equation}
with $M_\pm\equiv M_b\pm M_c$.
	
At the zero-recoil point of $\omega=1$, the form factors are simply written as~\cite{Falk:1992ws}
\begin{equation}\label{prec}
\begin{aligned}
&F^V_1(1)=1+\varepsilon_b+\varepsilon_c+\varepsilon_c^2(b_1-b_2)\,,&F_1^A=1+\varepsilon_c^2 b_1\,,&\\
&F_2^{V}(1)=F_2^A(1)=-\varepsilon_c+b_2\varepsilon^2_c,&F_{3}^{V,A}(1)=\mp\varepsilon_b\,,&
\end{aligned}
\end{equation}
to the precision of ${\cal O}(\varepsilon_b\varepsilon_c)$, providing that hard gluon corrections are absent. 
A great advantage in the heavy quark system is that the recoil effects are taken into account by a single function of $\overline{\xi}(\omega)$, i.e.
\begin{equation}
F_{1,2,3}^{V,A}(\omega) =\overline{\xi}(\omega) F_{1,2,3}^{V,A}(1) \,.
\end{equation}

\begin{figure}[h]
		\centering
		\includegraphics[width=0.95\linewidth]{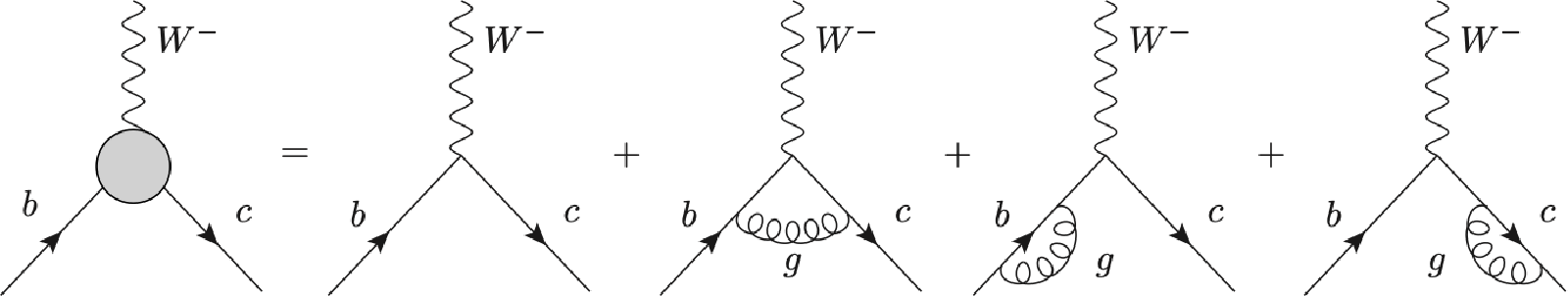}
		\caption{The hard gluon corrections of the current operators.}
		\label{hard}
	\end{figure}

	After including the hard gluon corrections shown in Fig.~\ref{hard}, the form factors receive several corrections, given by
	\begin{small}
		\begin{eqnarray}\label{gluon}
			&&\frac{F_1^V(\omega)}{\overline{\xi}(\omega) }=1+\hat{\alpha}_s C_{V_1}+\varepsilon_c+\varepsilon_b+\hat{\alpha}_s\left[C_{V_1}+2(\omega-1) C_{V_1}^{\prime}\right]\left(\varepsilon_c+\varepsilon_b\right)+\varepsilon_c^2\left(b_1-b_2\right)\,, \nonumber\\
			&&\frac{F_2^V(\omega)}{\overline{\xi}(\omega) } = \hat{\alpha}_s C_{V_2}-\frac{2 \varepsilon_c}{\omega+1}+\hat{\alpha}_s\left\{C_{V_2} \frac{3 \omega-1}{\omega+1} \varepsilon_b-\left[2 C_{V_1}-(\omega-1) C_{V_2}+2 C_{V_3}\right] \frac{\varepsilon_c}{\omega+1}\right.\nonumber\\
			&&\quad\quad\quad\quad\quad\quad\quad+2(\omega-1) C_{V_2}^{\prime}\left(\varepsilon_c+\varepsilon_b\right)\bigg\}+\varepsilon_c^2b_2\,,\nonumber\\
			&&\frac{F_3^V(\omega)}{\overline{\xi}(\omega) }=\hat{\alpha}_s C_{V_3}-\frac{2 \varepsilon_b}{\omega+1}+\hat{\alpha}_s\left\{ C_{V_3} \frac{3 \omega-1}{\omega+1} \varepsilon_c-\left[2 C_{V_1}+2 C_{V_2}-(\omega-1) C_{V_3}\right] \frac{\varepsilon_b}{\omega+1}\right.\nonumber\\
			&&\quad\quad\quad\quad\quad\quad\quad +2(\omega-1) C_{V_3}^{\prime}\left(\varepsilon_c+\varepsilon_b\right)\bigg\}\,,\\
			&&\frac{F_1^A(\omega)}{\overline{\xi}(\omega) }=1+\hat{\alpha}_s C_{A_1}+\left(\varepsilon_c+\varepsilon_b\right) \frac{\omega-1}{\omega+1}+\hat{\alpha}_s\left[C_{A_1} \frac{\omega-1}{\omega+1}+2(\omega-1) C_{A_1}^{\prime}\right]\left(\varepsilon_c+\varepsilon_b\right)+\varepsilon_c^2 b_1\,,\nonumber\\
			&&\frac{F_2^A(\omega)}{\overline{\xi}(\omega) }=\hat{\alpha}_s C_{A_2}-\frac{2 \varepsilon_c}{w+1}+\hat{\alpha}_s\left\{C_{A_2} \frac{3 w+1}{w+1} \varepsilon_b-\left[2 C_{A_1}-(w+1) C_{A_2}+2 C_{A_3}\right] \frac{\varepsilon_c}{w+1}\right.\nonumber\\
			&&\quad\quad\quad\quad\quad\quad\quad +2(w-1) C_{A_2}^{\prime}\left(\varepsilon_c+\varepsilon_b\right)\bigg\}+\varepsilon_c^2b_2\,,\nonumber\\
			&&\frac{F_3^A(\omega)}{\overline{\xi}(\omega) }=\hat{\alpha}_s C_{A_3}+\frac{2 \varepsilon_b}{w+1}+\hat{\alpha}_s\left\{C_{A_3} \frac{3 w+1}{w+1} \varepsilon_c+\left[2 C_{A_1}-2 C_{A_2}+(w+1) C_{A_3}\right] \frac{\varepsilon_b}{w+1}\right.\nonumber\\
			&&\quad\quad\quad\quad\quad\quad\quad +2(w-1) C_{A_3}^{\prime}\left(\varepsilon_c+\varepsilon_b\right)\bigg\}\,,\nonumber
	\end{eqnarray}\end{small}where $\hat{\alpha}_s$ is the strong coupling constant and the definitions of $C^{(\prime) }_{A,V_{1,2,3}}$
	can be found in Ref.~\cite{Bernlochner:2018kxh}. 
	To include both soft and hard gluon corrections, we extract the relevant parameters of $\overline{\xi}(\omega)$, $\overline{\Lambda}$, and $b_{1,2}$ from the quark models, where the soft gluon~(non-perturbative) effects are taken account by the wave functions. 
	After that, we plug  the computed parameters into Eq.~\eqref{gluon} to contain the hard gluon corrections. 
	
	We note that the heavy baryon masses can be expanded as
	\begin{equation}\label{MQ}
		M_Q=m_Q+\overline{\Lambda}+\frac{\lambda_1}{2m_Q}+{\cal O}(\varepsilon_Q^2).
	\end{equation}
	It is convenient to rewrite the binding energy as $\overline{\Lambda }=CE_{\text{di}}$, where $E_{\text{di}} $ stands for the energy of the diquark system, and $C$ describes the correction of it in the presence of the heavy quark with an infinite mass.
	
	\subsection{Homogeneous bag model}\label{sec2a}
	We begin our study with the MIT bag model, of which the quark wave functions of the baryons are confined in a finite region, given as~\cite{bag}
	\begin{equation}\label{quark_wave_function}
		\begin{aligned}
			&\phi_{q\updownarrow} (\vec{x}) = \left(
			\begin{array}{c}
				j_0(p_qr) \chi_\updownarrow\\
				i\frac{p_q }{E_q^k+p_q} j_1(p_qr) \hat{r} \cdot \vec{\sigma} \chi_\updownarrow\\
			\end{array}
			\right)\,,~~~~&\text{for}~r\leq R\,,\\
			&\phi_{q\updownarrow} (\vec{x})=0\,, &\text{for}~r>R\,,
		\end{aligned}
	\end{equation}
	with $r= |\vec{x}|$ and $R$ the bag radius. Here, $j_{0,1}$ are the spherical Bessel  functions,  $E_q^k = \sqrt{p_q^2 + m_q^2} $ with $m_q$ and $p_q$ the quark mass and three-momentum, and $\chi_\uparrow= (1,0)^T$ and $\chi_\downarrow= (0,1)^T$ represent a spin-up and a spin-down quarks, respectively.		
	From the boundary condition,
	$p_q$ must satisfy~\cite{bag}
	\begin{equation}\label{boundary}
		\tan(p_q R) = \frac{p_qR }{1 - m_qR + E_q^k R }\,.
	\end{equation}
	In particular, we have 
	\begin{equation}\label{lightlimit}
		\lim_{m_{u,d}\to 0 } p _{u,d}  = 2.043 /R \,,
	\end{equation}
	and 
	\begin{equation}\label{com2}
		p_{Q} = \frac{\pi }{R}\left(
		1 -\ \frac{1}{2 m_{Q}R }
		\right)\,,~~~E_{Q}^k = m_{Q}  + \frac{\pi^2 }{2m_{Q}R^2}\,,
	\end{equation}
	to the precision of ${\cal O}(1/m_{Q}^2R^2)$. It is interesting to point out that the zeroth order corrections are  absent in $E_Q^k$, and  the first order corrections correspond to  the kinematic energies of the heavy quarks in the nonrelativistic limit. Plugging Eq.~\eqref{com2} into the bag wave functions, we are led to
	\begin{equation}
		\phi_Q (\vec{x}) = \left(
		\begin{array}{c}
			\left(  	j_0(\pi r/ R) -\frac{\pi r}{2R^2m_Q}j_1(\pi r/R)\right) \chi\\
			i\frac{\pi }{2m_Q R} j_1(\pi r/R ) \hat{r} \cdot \vec{\sigma} \chi\\
		\end{array}
		\right)\,.
	\end{equation}

	On the other hand, the masses of the baryons are given as 
	\begin{equation}
		M_{Q} = Z_0/R+\frac{4\pi}{3}R^3 B_0^4 +E_I+  \sum_q E_q^k  \,,
	\end{equation}
	where $Z_0$ and $B_0$  are associated with the zero-point and bag volume energies, respectively, $E_I$ is the interaction energies, 
	and $R$ minimalizes the baryon masses, given by
	\begin{equation}
		\frac{\partial M_Q}{\partial R} = 0 \,.
	\end{equation}
	In calculating the Isgur-Wise function, we do not include the interaction corrections, so we set $E_I=0$ for consistency, which is the major source of errors.
	
	Combining Eqs. \eqref{lightlimit} and \eqref{com2}, 
	we arrive at 
	\begin{equation}
		\frac{\partial M_{Q }}{\partial R} = 4\pi R^2 B_0^4 - (Z_0 + 4.086 )/R^2=0\,.
	\end{equation}	
	There are two sets of bag parameters in Ref.~\cite{bag}, given as 
	\begin{equation}
		(Z_0,B_0)  = (1.84, 0.145~\text{GeV}) , ( 1.95 , 0.125~\text{GeV} )\,,
	\end{equation}
resulting in that 
	\begin{equation}\label{u3}
		(\overline{\Lambda}  , R^{-1}) = (0.665, 0.223) , (0.554, 0.195)~\text{GeV} ,
	\end{equation}
respectively.

	The twist is that although the MIT bag model  successfully explains most of the low-lying baryon masses, it is  difficult to be applied in decays,
	especially  in taking account the recoil effects~($\omega$ dependencies).
	The problem can be traced back to that bag states are localized. According to the the Heisenberg uncertainty principle, it cannot be an eigenstate of three-momenta, which is referred to as the center-of-mass motion~(CMM) problem. 
	The difficulties were tackled a few years ago in Ref.~\cite{Geng:2020ofy}, where localized bags  are  replaced by linear superpositions of infinite ones, distributed homogeneously all over the space. 

It has been shown that after the CMM is removed, the axial form factor in the neutron beta decay increases to $g_A=1.31$ by 20\%~\cite{Liu:2022pdk}. Comparing to the experimental value of $g_A=1.275$, it is clear that the numerical estimation is improved. On the other hand,   in the heavy flavor conserving decays, the CMM was identified to be the reason of the underestimation of the four-quark operator matrix elements~\cite{Cheng:2022jbr}. 
 
	In the HBM, the baryon states at rest  are
	given as 
	\begin{eqnarray}\label{wave}
		&&	|\Lambda_Q,\uparrow \rangle = \int\frac{1}{\sqrt{6} } \epsilon^{\alpha \beta \gamma} d _{a\alpha}^\dagger(\vec{x}_d) u_{b\beta}^\dagger(\vec{x}_u) Q_{c\gamma}^\dagger (\vec{x}_Q) \Psi^{abc}_{A_\uparrow(duQ)} (\vec{x}_d,\vec{x}_u,\vec{x}_Q) [d^3  \vec{x}] | 0\rangle\,,
	\end{eqnarray}
	where the Greek~(Latin) letters represent the color~(Dirac spinor) indices, $\epsilon$ stands for the totally anti-symmetric tensor,  
	$[d^3 \vec{x}]=d^3 \vec{x}_d d^3 \vec{x}_u d^3 \vec{x}_Q$, $q^\dagger$ is the creation operator of the quark, satisfying
	\begin{equation}\label{com}
		\{ q^\dagger_{a \alpha}(\vec{x}) , q_{b \beta}(\vec{x}\,') \} = \delta_{ab} \delta_{\alpha\beta} \delta^3 \left(
		\vec{x} - \vec{x} \,'
		\right)\,,
	\end{equation}
	with $q \in\{ u,d,Q\}$, and the spatial distributions are described by 
	\begin{equation} \label{xdelta}
		\begin{aligned}
			\Psi_{A_{\uparrow}(duQ)}^{abc} ( \vec{x}_d, \vec{x}_u ,  \vec{x}_Q ) =& \frac{{\cal N}_Q}{\sqrt{2}} \int \left [ \phi^a_{d\uparrow}(\vec{x}_d- \vec{x}_\Delta) \phi^b_{u\downarrow}(\vec{x}_u-\vec{ x}_\Delta) \right. \\
			&\left. - \phi^a_{d\downarrow}(\vec{x}_d- \vec{x}_\Delta) \phi^b_{u\uparrow}(\vec{x}_u-\vec{ x}_\Delta) \right]  \phi^c_{Q \uparrow}(\vec{x}_Q- \vec{x}_\Delta) d^3 \vec{x}_\Delta,
		\end{aligned}
	\end{equation}
	where ${\cal N}_Q$ are the normalization constants.
    If not stated otherwise, $q^\dagger (\vec{x})$ is evaluated at $t=0$ in this work.

	In Eq.~\eqref{xdelta},  $\phi_{q\updownarrow}(\vec{x}_q -\vec{x}_\Delta)$ represents the  quark state in the static bag centering at $\vec{x}_\Delta$. 
	By carrying out the integral of $d^3 \vec{x}_\Delta$, it is straightforward to see that the baryons are distributed uniformly all over the three-dimensional space. 
	Therefore, the unwanted CMM is removed. We note that
	the formulas are reduced to the ones of the MIT bag model 
	when the integral of $d^3\vec{x}_\Delta$ is eliminated.
	
	To get a baryon state with a nonzero momentum, 
	we apply the Lorentz boost to Eq.~\eqref{wave}. 
	The transformation rule of the creation operators reads
	\begin{equation}\label{Lorentz}
		U_v q^\dagger _{\alpha a} ( x^\mu  ) U_v^{-1} = \left( S_v \right) _{ab} q ^ \dagger _{ \alpha b} \left(\left( \Lambda_v^{-1}\right) ^\mu \,_\nu x^\nu\right ) \,,
	\end{equation}
	where $U_v$, $S_v$, and $\Lambda_v$ are the Lorentz boost operators of  states, Dirac spinors, and coordinates, respectively. Without lost of generality, we take the Lorentz boost toward the $\hat{z}$ direction, resulting in 
	\begin{equation}
		S_{\pm v}   = \left(
		\begin{array}{cc}
			a_+ & \pm a_-\sigma_3\\
			\pm a_-\sigma_3  & a _ +
		\end{array}
		\right)\,,~~~S_{\pm v} ^2  = \left(
		\begin{array}{cc}
			\gamma & \pm \gamma v\sigma_3\\
			\pm \gamma v\sigma_3  & \gamma 
		\end{array}
		\right)\,,
	\end{equation}
	where $a_\pm = \sqrt{(\gamma \pm 1)/2}$ and $\gamma = 1/\sqrt{1-v^2}$. 
	
	From 
	Eq.~\eqref{Lorentz}, we see  that even if we start with $t=0$, the $t$ dependencies of quark states are nevertheless still required.
	It is due to that the $t=0$ plane is not invariant under Lorentz boosts. Thus, for the state having a Lorentz boost, we have to evaluate the anticommutation relations with unequal times, which cannot be treated perturbatively.
	To overcome the problem, 
	we utilize  that the quark states are  energy eigenstates in the bag model, and make the substitution in  Eq.~\eqref{xdelta} 
	\begin{eqnarray}\label{ene}
		q^\dagger _{a\alpha}(\vec{x}) =  e^{-iE_qt} q^\dagger _{a\alpha}(t, \vec{x} ) \,,
	\end{eqnarray}
	where $E_q$ are the energies of the quarks. 
	It is clear that we do not consider the interaction energies here, as the quark energies are independent to each other.  
	The reasonable range of $E_q$ is   
	\begin{equation}\label{u1}
		\frac{1}{3} M_N <E_q 	< E_q^k + \frac{1}{3}(E_0+E_v) \,,
	\end{equation}
	where $M_N$ is the neutron mass, $E_0$ and $E_v$ are the zero-point  and  vacuum energies arisen from the bag,  respectively,
	which are allocated evenly among the quarks. 
	In the HBM, Eq.~\eqref{u1} serves as the major source of the uncertainties. 
	
After some algebra,
we arrive at~\cite{Liu:2022pdk}
\begin{equation}\label{boo}
\Psi ^{abc} (\vec{x}_1, \vec{x}_2, \vec{x}_3 ) \overset{v}{\to } (S_v)_{aa'}(S_v)_{bb'}(S_v)_{cc'} \Psi ^{a'b'c'}(\vec{x}_1^v, \vec{x}_2^v, \vec{x}_3^v ) ,
\end{equation}
where $\vec{x}^v = (x,y,\gamma z)$. 
 From the normalization condition
	\begin{equation}
		\langle \Lambda_Q , \vec{p}\,' | \Lambda_Q \vec{p} \,\rangle =  u^\dagger u (2\pi) ^3  \delta^3 (\vec{p}- \vec{p}') \,,
	\end{equation}
	we derive that 
	\begin{equation}\label{normal}
		\frac{\overline{u} u }{{\cal N}^2 }= \int d^3\vec{x}_\Delta \prod_{q}\int    \phi_{q}^\dagger \left(\vec{x}_q^{\,+}\right) \phi_{q} \left(\vec{x}_q^{\,-}\right) d^3\vec{x}_q\,,
	\end{equation}
	where $\vec{x}^\pm_q = \vec{x}_q \pm \vec{x}_\Delta /2 $, and 
	$\sum_q E_q = M_Q$ has been used. 
	
We adopt the Breit frame so that $\Lambda_b^0$ and $\Lambda_c^+$ have opposite velocities. Collecting Eqs.~\eqref{wave}, \eqref{com} and \eqref{boo}, 
we find that
	\begin{eqnarray}\label{master}
		\langle \Lambda_c^+ (\vec{v}\,) | c^ \dagger\Gamma  b (0)|\Lambda_b^0  (- \vec{v}\,) \rangle  = {\cal N}_c \,{\cal N}_b  \int d^3\vec{x}_\Delta \Gamma _{cb}(\vec{x}_\Delta) \prod_{l } D^v_{l}(\vec{x}_\Delta)\,,
	\end{eqnarray}
	along with
	\begin{eqnarray}\label{conmas}
		&&\Gamma _{cb} (\vec{ x}_\Delta)=\int  d^3\vec{x}  \phi _c^\dagger\left(\vec{x} + \frac{1}{2}\vec{x}_\Delta \right) S_{ v}\Gamma S_{- v} \phi_b\left(\vec{x} - \frac{1}{2}\vec{x}_\Delta \right) e^{2iE_{\text{di}}\vec{ v}\cdot \vec{ x}     }\,,\nonumber \\
		&&D^v_l (\vec{x}_\Delta ) = \frac{1}{\gamma}
		\int d^3 \vec{x } \phi_l^\dagger\left(\vec{x}+ \frac{1}{2} \vec{x}_\Delta \right)  \phi_l \left(\vec{x}- \frac{1}{2} \vec{x}_\Delta\right)  e ^ { - 2 i E_l\vec{v}\cdot\vec{x}} \,,~~~l=u,d\,,
	\end{eqnarray}
	where 
	$\Gamma $ is an  arbitrary  Dirac matrix,  $E_{\text{di}} = E_u+E_d$, and $\omega =\gamma^2(1+v^2)$.
	In general, $\Gamma_{cb}$ and $D_l^v$ would be  some complicated functions of the quark masses and bag radius. To examine the model, we take  $\vec{v}\to 0$ and  Taylor expand  the formulas regarding to  $M_Q$, leading to~\footnote{In the calculation of the form factors, we have taken the normalization 
	of $\overline{u}_Qu_Q = 1$.}
	\begin{equation}\label{com4}
		\frac{1}{{\cal N}_\infty ^2}
		= 16\pi^2\int \left(  D^0_{l} (r_\Delta )\right)^2 r_\Delta ^2 dr_\Delta  \int_0^{ \sqrt{R^2-r_\Delta^2/4}} d \rho \rho  \int_0^{ \sqrt{R^2-r_\Delta^2/4} - r_\Delta/2 }  d z 
		j_0^{+ } j_0^{- } \,,
	\end{equation}
	with the abbreviations of $j_{0,1}^\pm = j_{0,1}(\pi r_\pm /R )$, $r_\pm = |\vec{x} \pm \vec{x}_\Delta/2 |$ and $r_\Delta = |\vec{x}_\Delta|$. 
In addition, with $m_{u,d}\to 0$, we are led to 
	\begin{equation}
		D^0_l (r_\Delta) = 4\pi \int_0^{ \sqrt{R^2-r_\Delta^2/4}} d \rho \rho  \int_0^{ \sqrt{R^2-r_\Delta^2/4} - r_\Delta/2 }  d z \left[
		l_0^{+ } l_0^{- } + \left( r ^2 -  r_\Delta^2/4 \right) l_1^{+}l_1^{-}
		\right] \,,
	\end{equation}
where $l_{0,1}^{\pm} = j_{0,1} (2.043 r_\pm /R )$.
From Eq.~\eqref{conmas},
the integrals are suppressed by the oscillations of the exponential functions, which depend heavily on $E_q$. In the  case of $\Lambda_b^0 \to \Lambda^0 \gamma$,  the ambiguity in Eq.~\eqref{u1} causes the calculated branching fraction to vary from $3.5\times 10^{-6}$ to $1.0\times 10^{-5}$~\cite{Liu:2022pdk}. However, 
$E_q$ is always followed by $\vec{v}$, so its uncertainty does not affect the results at $\vec{v}=0$. In this case, $b_{1,2}$ are uncontaminated by the uncertainties of $E_q$ as they are evaluated at $\vec{v}=0$. 
 
The form factors of
	$F_1^{V}$ and $F_1^A$ are extracted by
	\begin{eqnarray}\label{com3} 
		&& F_1^{V} (1)  =  \lim_{v\to0}  \frac{1}{v}  \langle \Lambda_c^+ (\vec{v}\,) | \left( \overline{c} \,\gamma^1 b \right)  (0) |\Lambda_b^0  (- \vec{v}\,) \rangle \,,\nonumber \\
		&&F_1^{A} (1)  =  \lim_{v\to0}   \langle \Lambda_c^+ (\vec{v}\,) | \left( \overline{c} \,\gamma^1\gamma_5  b \right)  (0) |\Lambda_b^0  (- \vec{v}\,) \rangle,
	\end{eqnarray}
resulting in that 
	\begin{eqnarray}
		&&F_1^V (1)= 1 +\left( 
		\frac{1}{2m_c}+\frac{1}{2m_c}\right) E_{\text{di}} {\cal N}_\infty ^2
		4\pi \int r_\Delta^2  d r_\Delta
		{\cal C}(r _\Delta) 
		\left(  D^0_{l} (r_\Delta )\right)^2
		\,,\nonumber\\
		&&F_1^A (1)= 1\,,
	\end{eqnarray}
	where 
	\begin{eqnarray}\label{ress0}
		&&{\cal C} (r_\Delta)  = \frac{4 \pi^2 }{3 R} \int_0^{ \sqrt{R^2-r_\Delta^2/4}} d \rho \rho  \int_0^{ \sqrt{R^2-r_\Delta^2/4} - r_\Delta/2 }  d z \Big[  \left(
		j_0^+ j_1^-\frac{ 1}{r_-}+ j_1^+ j_0 ^- \frac{ 1}{r_+}
		\right) r^2\nonumber\\
		&&+\left(
		j_0^- j_1^+\frac{ r_\Delta z}{2r_+} - j_1^- j_0 ^+ \frac{ r_\Delta z }{2 r_- }
		\right) 	\Big]\,,
	\end{eqnarray}
	to the precision of ${\cal O}(1 / m_c^2R^2)$. By matching it  with  Eq.~\eqref{prec}\,, 
	we find that the first order correction of $F_1^A$ indeed vanishes,  and 
	\begin{equation}\label{ress}
		C=  {\cal N}_\infty ^2
		4\pi \int r_\Delta^2  d r_\Delta
		{\cal C}(r _\Delta) 
		\left(  D^0_{l} (r_\Delta )\right)^2
		=  1\,, 		\end{equation}
	which holds exactly by an actual calculation. 
It is a sensible result, since  the zeroth order corrections of $E_Q^k$ are absent as shown in Eq.~\eqref{com2}, and the interactions among quarks are omitted.  
We conclude that the diquark energies are unaffected by the presence of the heavy quark at the zeroth order, namely $C=1$. However, Eq.~\eqref{ress} itself is a nontrivial result, and it indicates that our treatments for the CMM are self-consistent.	
	
We emphasize that  Eq.~\eqref{ress} is a parameter-independent result, which  can  be shown explicitly by changing the variables 
	\begin{equation}
		(\rho, z, \vec{x}_\Delta)  \to \left(\frac{\rho}{R}, \frac{z}{R}, \frac{\vec{x}_\Delta}{R}\right)  \,,
	\end{equation}  
in Eqs.~\eqref{ress0} and \eqref{ress}.
	To take account the recoil effects, we use the following trick
	\begin{eqnarray}
		&&\xi(\omega) = \left. \sum_{i=1,2,3}F_i^V(\omega) \right|_{\varepsilon_Q =0}\,,\nonumber\\
		&& \xi_{ke}(\omega)= 
		\left. \frac{\partial}{\partial \varepsilon_c}\left(
		\sum_{i=1,2,3}F_i^V(\omega) \right) \right|_{\varepsilon_Q=0}\,,
	\end{eqnarray}
	where $\sum_i F_i^V$  is evaluated by taking $\Gamma = 1$ in Eq.~\eqref{conmas}. Consequently, we get
	\begin{eqnarray}
		\xi(\omega) &=& 
		{\cal N}_\infty ^2
		\int d^3 \vec{x}_\Delta  d^3\vec{x }
		\left(
		j_0^+ j _0^-   
		\right) e^{2iE_{\text{di}}\vec{v}\cdot\vec{x}}\left(  D^0_{l} (\vec{x}_\Delta )\right)^2
		\,,\\
		\xi_{ke}(\omega) &=& A(\vec{v}\,) \overline{\Lambda} - A(0)\overline{\Lambda}
		\nonumber\,,\\
		A(\vec{v}) &\equiv& \frac{\pi {\cal N}_\infty ^2 }{ R^2  }
		\int d^3 \vec{x}_\Delta d^3\vec{x }
		\left(r_-   j_0^+ j_1 ^- + r_+ j_1^+ j_0^- \right ) e^{2iE_{\text{di}}\vec{v}\cdot\vec{x}}
		\left(  D^0_{l} (\vec{x}_\Delta )\right)^2\,.
		\nonumber
	\end{eqnarray}
	By taking $\vec{v}= 0$, we get that $\xi(1) = 1$ and $\xi_{ke}(1) =0$ as demanded by the Luke's theorem. 
	
	Similarly, the first order corrections of $b_2$ and $b_2$ are obtained, given by
	\begin{eqnarray}
		&&\left.\frac{\partial^2 F_1^V(1)}{\partial \varepsilon_c^2}\right|_{\varepsilon_Q =0} = (b_1 - b_2) \,,\nonumber\\
		&&\left.\frac{\partial^2 F_1^{A}(1)}{\partial \varepsilon_c^2}\right|_{\varepsilon_Q =0} = b_1\,, \quad
		\left.\frac{\partial^2 F_2^{V,A}(1)}{\partial \varepsilon_c^2}\right|_{\varepsilon_Q =0} = b_2\,.
	\end{eqnarray}
	The operations can be easily done by a computer program. However, their expressions are much more complicated and lengthy as well, so we do not list  them out here. As a cross-check, we compute
	\begin{equation}
		\left.\frac{\partial^2 }{\partial \varepsilon_c^2}
		\left(F_1^V (\omega)-F_1^A(\omega) + F_2^{V,A}(\omega)
		\right)
		\right|_{\varepsilon_Q =0}\,,
	\end{equation}
	which is indeed found to be zero, consistent with Eq.~\eqref{prec}.  
	
We note that  $\xi$, $\xi_{ke}$ and
$b_{1,2}$ depend only on $E_{\text{di}} R$, which is the only dimensionless parameter. 
To be specific, they are invariant under the transformation of $(R,E_{\text{di}})\to (a^{-1}R,aE_{\text{di}})$, where $a$ is an arbitrary constant. 
For a practical purpose, we can fix $R$ and vary $E_{\text{di}}$ solely  to cover the model uncertainties on $\overline{\Lambda}$, $\xi$, $\xi_{ke}$ and
$b_{1,2}$. 
	
	\subsection{Light-front quark model}
	
	In the LFQM,
	the baryon states are expressed as
	\begin{equation}\label{43}
		\begin{aligned}
			|\Lambda_Q,\uparrow\rangle=&\int\frac{1}{\sqrt{6}}\epsilon^{\alpha\beta\gamma}u^\dagger_{\alpha}(\tilde{p}_u,\lambda_u)d^\dagger_{\beta}(\tilde{p}_d,\lambda_d)Q^\dagger_{\gamma}(\tilde{p}_Q,\lambda_Q)\Psi^{[\lambda]}_{Q}(\tilde{p}_u,\tilde{p}_d,\tilde{p}_Q)[d^{3}\tilde{p}]|0\rangle,
		\end{aligned}
	\end{equation}
	where $\Psi^{[\lambda]}_{Q}(\tilde{p}_u,\tilde{p}_d,\tilde{p}_Q)$ represent the vertex functions between the $\Lambda_Q$ and $udQ$, $[\lambda]$ and  $[d^{3}\tilde{p}]$ stand for the light-front helicities $(\lambda_u,\lambda_d,\lambda_Q)$ and  light-front three-momentum integrals $d^{3}\tilde{p}_ud^{3}\tilde{p}_dd^{3}\tilde{p}_Q$, respectively, and 
	\begin{equation}
		p_q=(p_q^-,\tilde{p}_q)=(p_q^-,p_q^+,p_{q\perp}),~d^{3}\tilde{p}_{q}\equiv \frac{d p_{q}^{+} d^{2} p_{q \perp}}{2(2\pi)^{3}},
	\end{equation}
	with $p_q^{\pm}=p_i^0\pm p_q^3$ and $p_{q\perp}=(p_q^1\,,p_q^2)\,$.
At the equal light-front time~($x^+ = 0$),	the commutation relationships for $q^\dagger_\alpha(\tilde{p}_q,\lambda_q)$ are 
	\begin{equation}
		\{q^\dagger_\alpha(\tilde{p},\lambda),q_\beta(\tilde{p}^\prime,\lambda^\prime)\}=\delta_{\lambda\lambda^\prime}\delta_{\alpha\beta}\delta^3(\tilde{p}-\tilde{p}^\prime).
	\end{equation}
	Notice that the wave functions are made of  position eigenstates in the HBM shown in Eq.~\eqref{wave},  whereas they are built out of momentum eigenstates here in contrast. 
	
	The vertex functions are further decomposed as
	\begin{equation}\label{46}
		\Psi^{[\lambda]}_{Q}=2(2\pi)^{3}\frac{1}{\sqrt{P^{+}}}\delta^{3}\left(\tilde{P}-\tilde{p}_u-\tilde{p}_d-\tilde{p}_Q\right)\Phi(\tilde{p}_u,\tilde{p}_d,\tilde{p}_Q)\Xi^{1/2,\uparrow}(\lambda_u,\lambda_d,\lambda_Q),
	\end{equation}
where $ \Phi $ are the momentum distribution functions of the quarks,  
	$\Xi^{1/2,\uparrow}$ stands for the helicity wave function
	\begin{equation}
		\Xi^{1/2,\uparrow}(\lambda_u,\lambda_d,\lambda_Q)=\sum_{s_u, s_d, s_Q}\prod_{q}\langle\lambda_{q}|R_{q}^{\dagger}|s_{q}\rangle \langle\frac{1}{2} s_{u}, \frac{1}{2} s_{d}, \frac{1}{2} s_{Q }\mid S S_{z}\rangle,
	\end{equation}
and $ R_q $ is the Melosh matrix, which brings the quark from its spin state at rest to a light-front helicity state with momentum $\tilde{p}_q$~\cite{Chiu:2017ycx}.
	
In this work, we consider the two-particle forces between the quarks, which are effectively described by  the harmonic oscillator potentials 
\begin{equation}
	V=\beta^4r^2/(2M),
\end{equation}
where $r$ are the distances of the quarks, and $M$ and $\beta$ are the reduced masses and  shape parameters, respectively, to be specified later.
We take the quarks of $u$ and $d$ to form a diquark cluster denoted as  $[ud]$, as they are (iso)spin singlet.
By integrating out the delta functions $\delta(\tilde{P} -\tilde{p}_u-\tilde{p}_d-\tilde{p}_Q)$,
Eqs.~\eqref{43} and \eqref{46} can be rewritten as
\begin{equation}
[d^3\tilde{p}] \to d^3\vec{q}\,d^3\vec{Q}\,,~~~\Phi(\tilde{p}_u,\tilde{p}_d,\tilde{p}_Q)\to\Phi(\vec{q},\vec{Q})
\end{equation}
where
\begin{equation}
	\Phi(\vec{q},\vec{Q})=\phi_{ud}\,\phi_{Q[ud]}=(\pi\beta_{ud}\beta_{Q[ud]})^{-3/2}\exp(-\frac{\vec{q}\,^{2}}{2\beta_{ud}^2}-\frac{\vec{Q}^2}{2\beta_{Q[ud]}^2}),
\end{equation}
$ \phi_{ud} $ and $\phi_{Q[ud]}$ are the wave functions of $(u,d)$ and $(Q,[ud])$,  and the shape parameters of $\beta_{ud}$ and $\beta_{Q[ud]}$  are  the typical relative three-momenta, respectively.
Note that we have projected the light-front three-momenta to the instant forms by defining
	\begin{equation}
		\begin{aligned}
			&x_Q\equiv\frac{p_Q^+}{P^+}=\frac{E_Q-Q_z}{E_{ud}+E_Q},~&Q_\perp=x_Q(p_{u\perp}+p_{d\perp})-(1-x_Q)p_{Q\perp}&,\\
			&y\equiv\frac{p_u^+}{p_u^++p_d^+}=\frac{E_d-q_z}{E_u+E_d},~&q_\perp=yp_{u\perp}-(1-y)p_{d\perp}&,
		\end{aligned}
	\end{equation}
where $E_{u,d}=\sqrt{\vec{q}\,^2+m^2}$, $E_Q=\sqrt{\vec{Q}^2+m_Q^2}$, $E_{ud}=\sqrt{\vec{Q}^2+(E_u+E_d)^2}$, and $m$ stands for the constituent quark masses of $u$ and $d$ in the LFQM. 
	
	By utilizing that the potentials are independent of quark masses, we find
	\begin{equation}\label{tempting}
		\begin{aligned}
			\beta_{Q[ud]}^4M_{Q[ud]}^{-1}=\beta_{d[ud]}^4M_{d[ud]}^{-1}=\beta_{ud}^4(m/2)^{-1}
		\end{aligned}
	\end{equation}
where $M_{q[ud]}$ is the reduced effective mass of $q$ and $[ud]$, given as 
	\begin{equation}
		M_{q[ud]}^{-1}=m_q^{-1}+E_{\text{di}}^{-1}.
	\end{equation}
	The second equality in Eq.~\eqref{tempting} comes from the Isgur-Karl model for the equal mass scenario~\cite{Isgur:1978xj}.
	By considering the well-measured ratio of $g_A/g_V=-1.275$ from the neutron beta decay, we get 
\begin{equation}\label{neutron}
\beta_{d[ud]}=0.952 m\,.
\end{equation}
To the precision of ${\cal O}(m_Q^{-2})$, we have 
	\begin{equation}\label{binf}
		\beta_{Q[ud]}=\left(1-\frac{E_{\text{di}}}{4m_Q}+{\cal O}(\frac{1}{m_Q^2})\right)\beta_{\infty},~~~\beta_{\infty}=(1+E_{\text{di}}/m)^{1/4}\beta_{d[ud]}.
	\end{equation}
The existence of the heavy quark limit for $\beta_\infty$ has already been discussed and put by hand in Refs.~\cite{Cheng:2004cc,Ke:2007tg}. Here, we provide a clear theoretical background to justify their assumptions.

Now, we are ready to compute $\overline{\Lambda},b_{1,2},\xi(\omega)$, and $\xi_{ke}(\omega)$.
We choose the frame of $q^+=0$ to avoid the $Z$-graph contribution~\cite{Choi:1998nf}.
We  set $m_b=m_c / x = m_Q$. At $q^2=0$,  $\omega$  corresponds to 
\begin{equation}
\omega(x) =\frac{M_Q^2+M_{Qx}^2}{2M_QM_{Qx}},
\end{equation}
with
\begin{equation}
M_{Q{(x)}} = (x) m_Q + \overline{\Lambda} + \frac{\lambda_1}{(x) 2  m_Q}
\end{equation}
from Eq.~\eqref{MQ}, leading to~\footnote{The form factors in the $q^+=0$ frame are extracted as Eq.~(24) in Ref.~\cite{Ke:2007tg}.}
\begin{equation}
    \begin{aligned}
    \xi(\omega(x))&=f_1(q^2=0,x,\varepsilon_Q\to0),\\
    \left(\frac{1}{x}+1\right)\xi_{ke}(\omega(x))&=\frac{\partial}{\partial\varepsilon_Q}f_1(q^2=0,x,\varepsilon_Q\to0).
		\end{aligned}
	\end{equation}
In reality, we have that $x\approx3$ and $\lambda_1=-0.5$~GeV$^{2}$, corresponding to $\omega = 1.43$. 
Since $q^2 = q^+ q^- - q_\perp^2$, the region of $q^2>0$ is polluted by the $Z$-graph contribution.
Nevertheless,  we utilize the fact that  $\xi(\omega)$ and $\xi_{ke}(\omega)$ are independent of $m_{b,c}$ for a fix $\omega$. In practice,
we can obtain the $\omega$ dependencies of  $\xi(\omega)$ and $\xi_{ke}(\omega)$ by varing $x$. 
	
For $\overline{\Lambda}$ and $b_{1,2}$, we again use the trick that they are independent of $m_{b,c}$.
We compute them with $x=1$. Accordingly, $g_1$ and $f_2$ are expanded as
	\begin{equation}\label{fgexp}
		g_1=1+\frac{g_1^{(2)}}{m_Q^2}+{\cal O}(1/m_Q^3),~f_2=\frac{f_2^{(1)}}{m_Q}+\frac{f_2^{(2)}}{m_Q^2}+{\cal O}(1/m_Q^3),
	\end{equation}
where
	\begin{equation}\label{g1f2}
		\begin{aligned}
			&g_1^{(2)}=-\frac{\beta_\infty^2}{2},\quad\quad f_2^{(1)}=\int d^3\vec{q}\,d^3\vec{Q}\,|\Phi_\infty|^2\frac{E_{ud}}{2},\\
			&f_2^{(2)}=\int d^3\vec{q}\,d^3\vec{Q}\,|\Phi_\infty|^2\frac{E_{ud}}{8\beta_\infty^2}\left[7E_{\text{di}}\beta_\infty^2-4E_{ud}\beta_\infty^2-2E_{\text{di}}Q^2\right],\\
		\end{aligned}
	\end{equation}
and $\Phi_\infty$ is the momentum wave function under the heavy quark limit, given as
    \begin{equation}\label{61}
	\Phi_\infty(\vec{q},\vec{Q})=(\pi\beta_{ud}\beta_\infty)^{-3/2}\exp(-\frac{\vec{q}\,^{2}}{2\beta_{ud}^2}-\frac{\vec{Q}^2}{2\beta_\infty^2}).
	\end{equation}
Note that the zeroth order of $g_1$ is consistent with the heavy quark symmetry, which is a nontrivial result.

To match Eqs.~\eqref{Ff} and \eqref{prec} to Eqs.~\eqref{fgexp} and \eqref{g1f2} with $x=1$,  there would be extra crossing terms at the order of $\varepsilon_c\varepsilon_b$ in Eq.~\eqref{prec}, which introduce two additional free parameters.
To eliminate them, we assume that $F^{V,A}_1(\omega)$  are factorized as
\begin{equation}
	F^{V(A)}_1(\varepsilon_b,\varepsilon_c, \omega)=f^{V(A)}(\varepsilon_c,\omega)f^{V(A)}(\varepsilon_b,\omega),
\end{equation}
with $f^{V,A}(\varepsilon_Q,\omega)$ to be determined.
Now we have
\begin{equation}\label{Lb1b2}
	\overline{\Lambda}=2f_2^{(1)},~\overline{b}_1=2g_1^{(2)},~\overline{b}_2=2(f_2^{(1)})^2-4f_2^{(2)}+2g_1^{(2)},
\end{equation}
where $\overline{b}_{1,2}\equiv b_{1,2}\overline{\Lambda}^2$.

From Eqs.~\eqref{tempting}, \eqref{g1f2} and \eqref{61}, we see that $f_2^{(1)}$ depends only on $E_{\text{di}}$ and $m$.
We arrive at 
\begin{equation}
f_2^{(1)}(E_{\text{di}}, m )=E_{\text{di}}{\cal F}\left(E_{\text{di}}, m\right)=E_{\text{di}}{\cal F}\left(\frac{E_{\text{di}}}{m}\right)\,,
\end{equation}
where ${\cal F}$ is a function to be determined, and the second equality comes from that ${\cal F}$ can only depends on the dimensionless quantities, {\it i.e.} $E_{\text{di}}/m$ in this case. 
By demanding the heavy quark limit of $\overline{\Lambda}=E_{\text{di}}$ or equivalently ${\cal F}=1/2$, we find 
\begin{equation}\label{Edi}
E_{\text{di}}=3.293 m\,.
\end{equation}
Collecting Eqs.~\eqref{tempting}, \eqref{g1f2}, \eqref{61} and \eqref{Edi},  we see that  only $m$ remains unfixed. 

In the LFQM, the HQE parameters depend on $m$, $E_{\text{di}}, \beta_\infty$
\begin{equation}\label{simpl}
h(m,E_{\text{di}}, \beta_\infty) =h(m,3.293m,0.952 m) = h(m)  \,, 
\end{equation}
where  $h \in \{  \xi(\omega), \xi_{ke}(\omega), b_1, b_2\}$, and we have used 
Eqs.~\eqref{neutron} and \eqref{Edi}. 

From the dimensional analysis, we find that  
\begin{equation}
h(m) = \sum_{n=-\infty}^\infty h'_n m^n\,,
\end{equation}
where $h'_n$ has $-n$ mass dimension. However, there is no other parameter with mass dimension, so we must have  $h'_n= 0 $ for $n\neq 0$. 

\begin{figure}[htbp]
		\centering
		\includegraphics[width=0.45\linewidth]{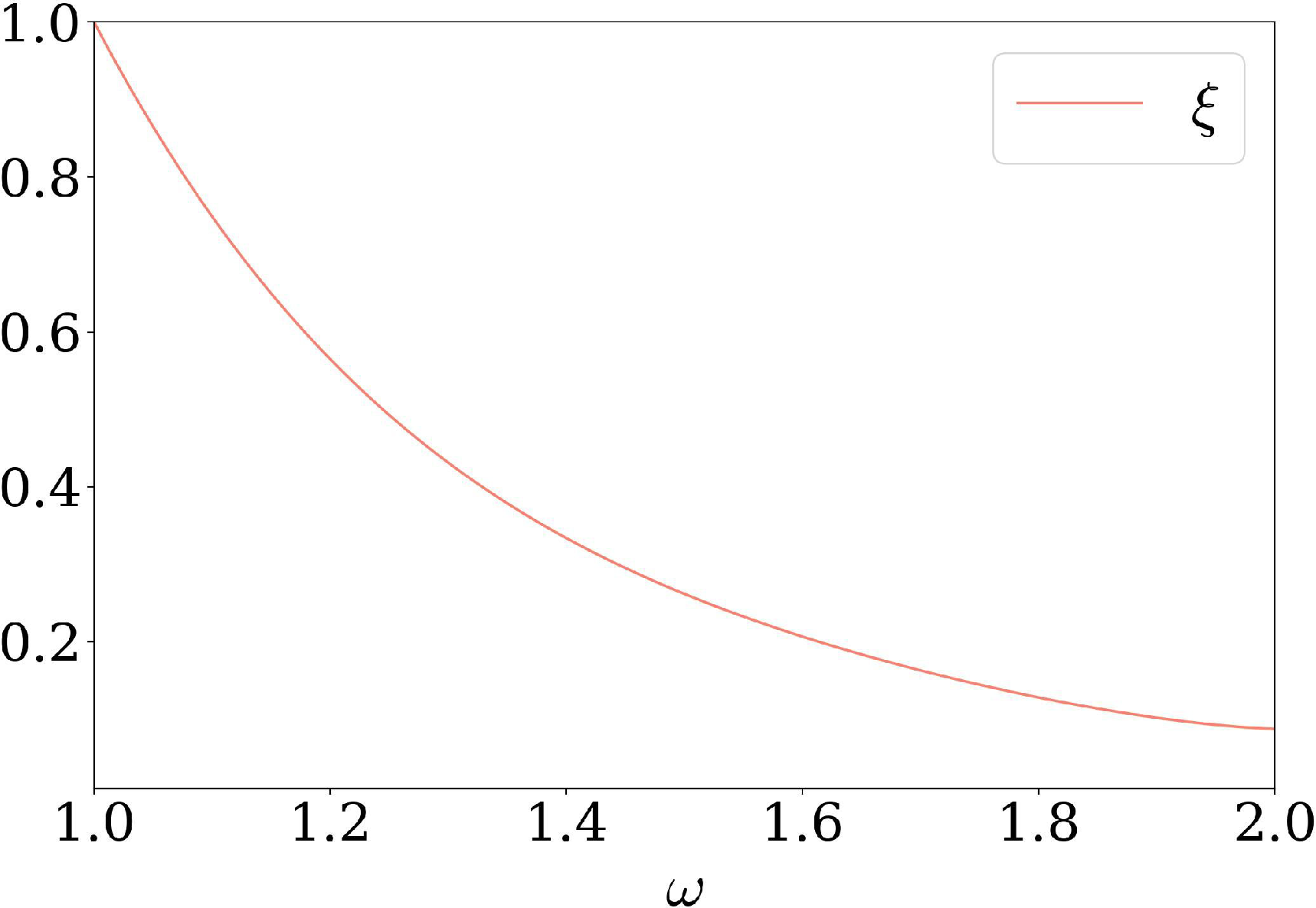}
		\includegraphics[width=0.45\linewidth]{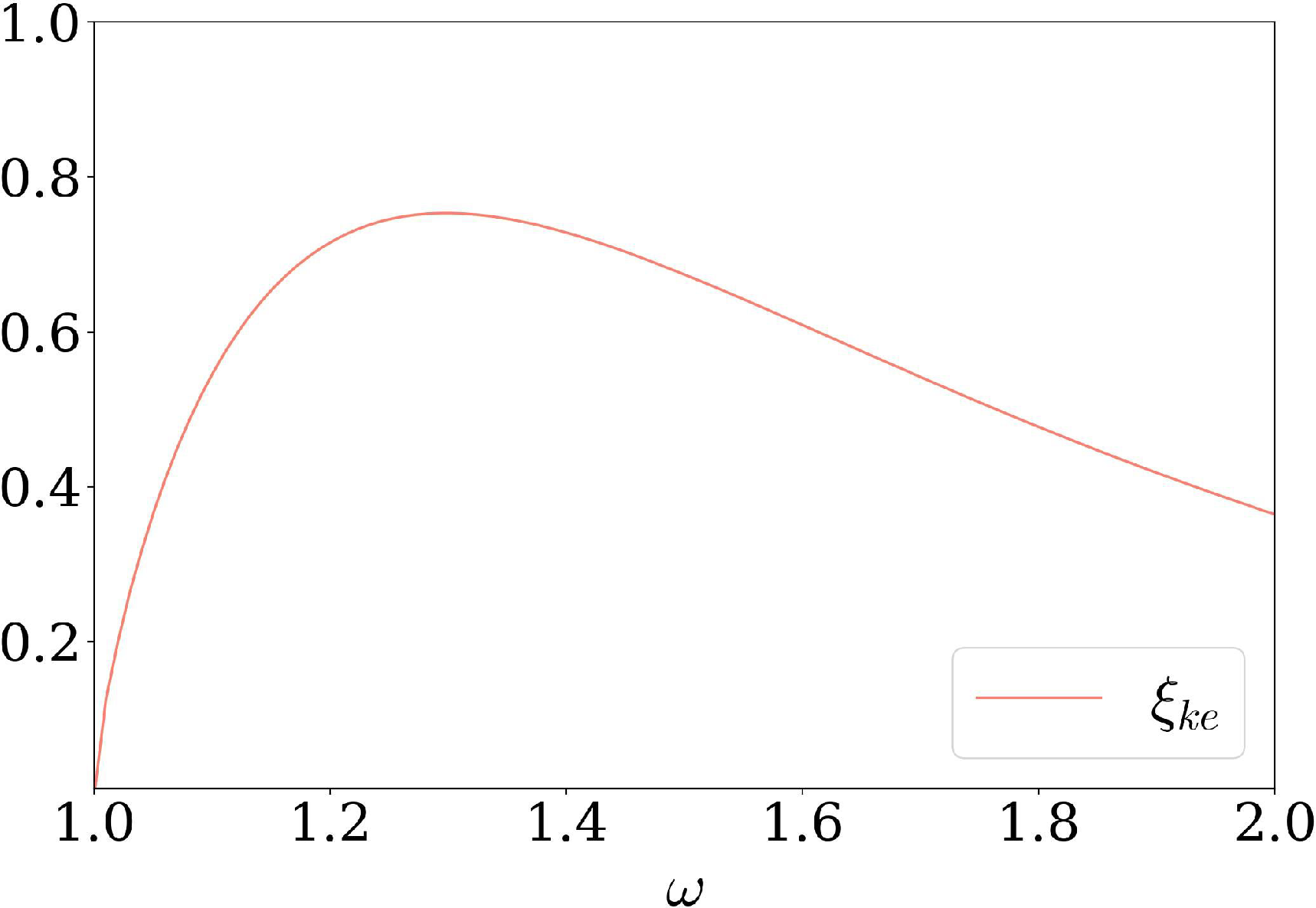}
		\caption{The Isgure-Wise function and its first order correction from the LFQM.}
		\label{XILF}
	\end{figure}

Consequently, $h$ do not depend on $m$ and are parameter-{\it independent} results in our approach. From Eqs.~\eqref{g1f2} and \eqref{Lb1b2}, we have 
\begin{equation}
b_1=-0.173,~b_2=0.518,
\end{equation}
while $\xi$ and $\xi_{ke}$ are shown in Fig.~\ref{XILF}. By a similar argument, we find $\overline{\Lambda}\propto m$.

\section{Numerical results}\label{sec3}

In both of the models,  $m_{b,c}$ are taken as the pole quark masses with $m_{b,c}= (4.78, 1.655)$~GeV, and 
the adopted value of the Cabibbo-Kobayashi-Maskawa~(CKM) matrix element is $|V_{cb}|=(42.2\pm0.7)\times10^{-3}$ from the inclusive semileptonic $B$ decays~\cite{pdg}. The formalism of the decay widths can be found in Ref.~\cite{Geng:2022hmf}.

For the HBM, we fix $R=4.8$~GeV$^{-1}$ without lost of generality as explained in the end of Sec.~\ref{sec2a}. From Eq.~\eqref{u1}, the reasonable range of $E_{\text{di}}$ is then given as 
	\begin{equation}
		0.585~\text{GeV}<E_{\text{di}}<0.686~\text{GeV}\,,
	\end{equation}
	which is consistent with Eq.~\eqref{u3}. 
For the LFQM, we take $m=0.30\pm0.08$ GeV, which is consistent with the experiments of the nucleon magnetic moments and ${\cal B}(\Lambda_b^0\to\Lambda^0\gamma)=(7.1\pm1.7)\times10^{-6}$~\cite{pdg,LHCb:2019wwi}. To be conservative, we allow $m$ to vary in a wide range, which shall cover all the reasonable values.
	
\begin{figure}
	\centering
	\includegraphics[width=0.45\linewidth]{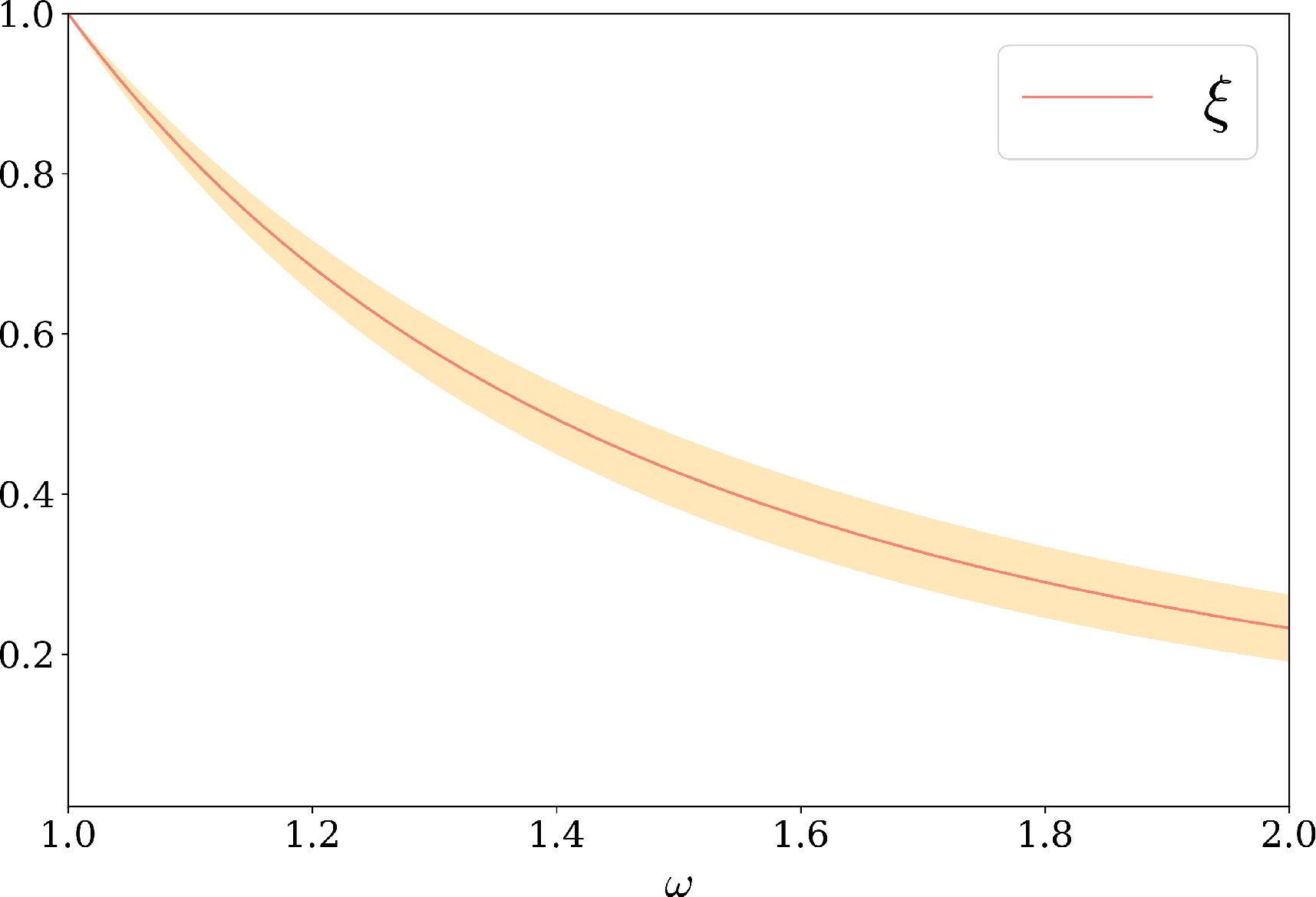}
	\includegraphics[width=0.45\linewidth]{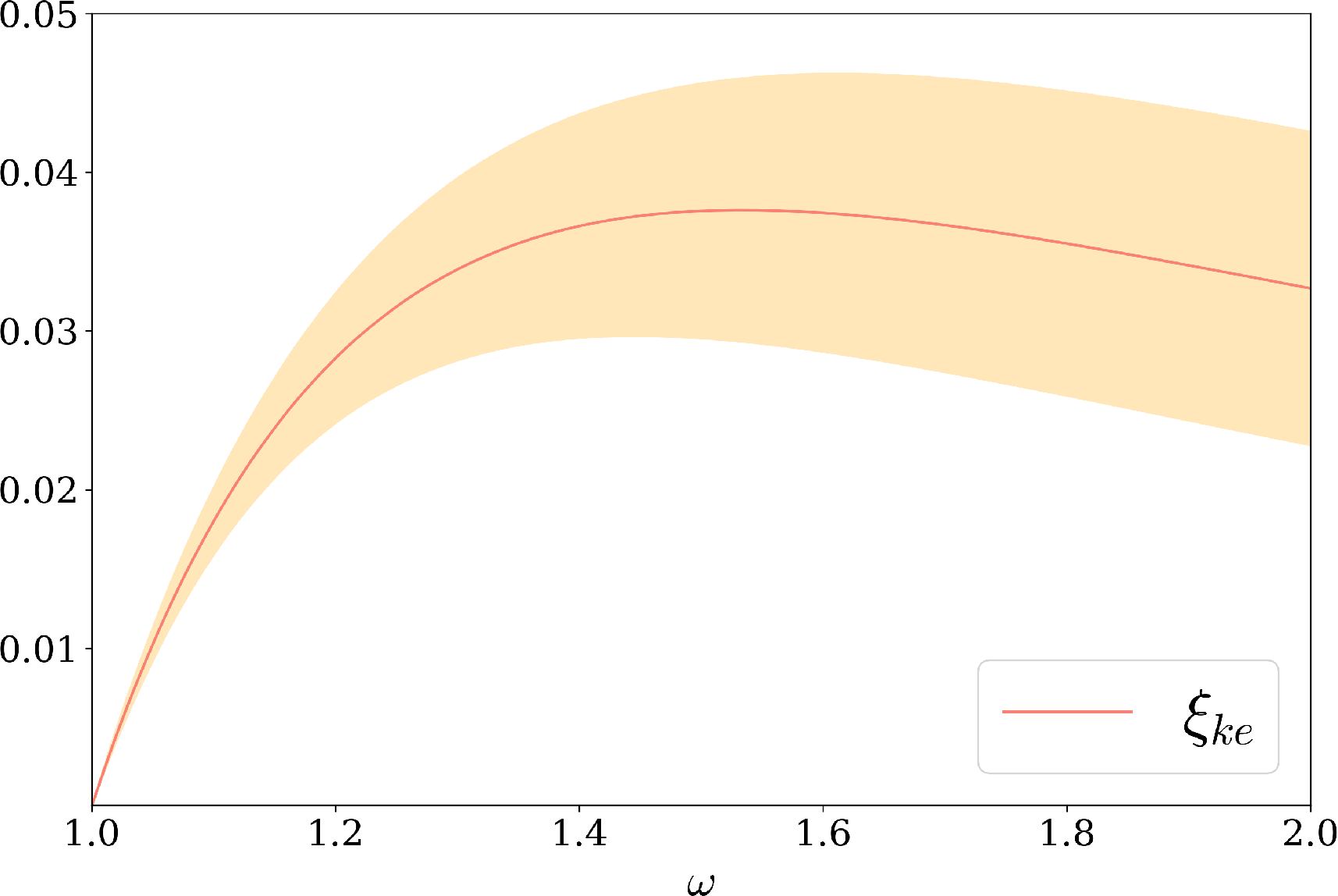}
	\caption{The Isgure-Wise function and its first order correction from the HBM.}
	\label{XI}
\end{figure}

The HQE parameters are computed with the formula given in Sec.~\ref{sec2}. The Isgur-Wise function and its first order correction of the HBM are given in Fig.~\ref{XI}.
Notice that the region of $\omega >1.43$ is equivalent to $q^2 <0$, and thus, it is irrelevant to $\Lambda_b^0 \to \Lambda_c^+\ell^-\overline{\nu}_\ell$.
Nevertheless, we have plotted them in the figures to see the  dependencies in the high $\omega$ region, as they are physical in the scattering processes.

To compare the results with those in the literature,
we expand the Isgur-Wise function regarding to $\omega$ to the fourth order, given as
\begin{equation}
		\overline{\xi}(\omega)=1+\overline{\xi}^{(1)}(\omega-1)+\frac{1}{2}\overline{\xi}^{(2)}(\omega-1)^2+\frac{1}{3!}\overline{\xi}^{(3)}(\omega-1)^3+\frac{1}{4!}\overline{\xi}^{(4)}(\omega-1)^4.
\end{equation}
The numerical results along with those in the literature are collected in Table~\ref{IW}. In the literature, Ref.~\cite{Bernlochner:2018kxh} fits the HQE parameters from the experimental data~\cite{LHCb:2017vhq} and LQCD calculations~\cite{Detmold:2015aaa}, Ref.~\cite{Ebert:2006rp} adopts the relativistic quark model~(RQM), Refs.~\cite{Miao:2022bga,Duan:2022uzm} employ the light-cone sum rule~(LCSR), and Ref.~\cite{Huang:2005mea} utilizes the QCD sum rule~(QCDSR).

We note that the authors of Ref.~\cite{Li:2021qod} have also considered the LFQM but with a different theoretical setup. 
On the one hand, their baryon wave functions are fitted from the mass spectroscopy with 13 free parameters, whereas ours are  based on the simple harmonic potential and HQE with one free parameter only.
On the other hand,  their form factors and Isgur-Wise function in the timelike region are obtained by the analytical continuation, where the dipole behavior is assumed. 
In our approach, we have mapped the dependency of $\omega$ to $x= m_{c}/m_{b}$, so that the Isgur-Wise function is directly evaluated in the entire phase space without further ad hoc assumptions. To sum up, we have shown that after considering the HQE, not only the parameter space of the LFQM is tightly constraint but also the ad hoc assumption of the analytical continuation is no longer needed. 

\begin{table}
\caption{The HQE parameters with $\overline{\Lambda}$ and $\overline{b}_{1,2}$ in units of GeV and GeV$^2$, respectively.}
\label{IW}
	\begin{tabular}{cccccccccc}
		\hline
		\hline			&HBM&LFQM&LHCb~\cite{LHCb:2017vhq,Bernlochner:2018kxh}&LQCD+LHCb~\cite{Bernlochner:2018kxh}&RQM~\cite{Ebert:2006rp}&QCDSR~\cite{Huang:2005mea}&LFQM~\cite{Li:2021qod}\\
		\hline
        \multirow{2}{*}{}$\overline{\xi}^{(1)}$&$-1.94(26)$&$-2.35(10)$&$-2.17(26)$&$-2.04(8)$&$-1.51$&$ -1.35(13) $&$1.67(11)$\\
        &&&&&&&$1.85(11)$\\
        \multirow{2}{*}{}$\overline{\xi}^{(2)}$&$4.78(93)$&$5.75(55)$&$ 4.10(105)$&$3.16(38)$&$4.06$&&$2.45(63)$\\
        &&&&&&&$3.25(61)$\\
        $\overline{\xi}^{(3)}$&$-10.8(25)$&$-10.5(21)$\\
		$\overline{\xi}^{(4)}$&$13.1(32)$&$10.1(38)$\\
		$\overline{\xi}(1.43)$&$0.470(43)$&$0.392(15)$&$0.446(4)$&$0.415(1)$\\
		\hline
		$\overline{\Lambda} $~~&$0.681(55)$&$0.988(263)$&$0.81(5)$\footnote{Extracted from $M_{\Lambda_Q}$, where $0.81(5)$ means $0.81\pm0.05$}&$0.81(5)^a$&$0.764$\\
	    $\overline{b}_1$&$-0.141(1)$&$-0.181(90)$&$0.24(192)$&$-0.46(15)$\\
    	$\overline{b}_2$&$0.351(35)$&$0.541(270)$&$0.45(188)$&$-0.39(39)$\\
		\hline
		\hline
	\end{tabular}
\end{table}

Surprisingly, the results in the HBM and LFQM agree well with each other, even though they are two very different quark models.
Our values of $\overline{\xi}^{(1)}$ also agree well with those from the   LHCb and LQCD, but significantly larger than the results from the RQM and  QCDSR.
Note that in Ref.~\cite{Bernlochner:2018kxh}, $\overline{\xi}^{(3)}$ and $\overline{\xi}^{(4)}$ have been omitted.
On the contrary, we find that they are sizable. In particular, $\overline{\xi}^{(3)}$ is  opposite in sign and  twice larger in comparison to  $\overline{\xi}^{(2)}$.
Although our results of $\overline{\xi}^{(2)}$ are larger than the values from the LHCb and LQCD, those of $\xi(\omega=1.43)$ are consistent with them  due to the presences of $\overline{\xi}^{(3)}$.

Note that our values of $\overline{b}_{1,2}$ from the LFQM are contaminated by the uncertainties of $\overline{\Lambda}$, but their signs are  not. Explicitly, both models give $\overline{b}_1<0$ and $\overline{b}_2>0$. The predicted sign of $\overline{b}_{2}$ is opposite to the one of LHCb+LQCD.
As the HBM and LFQM show  well consistence in all the HQE parameters, we are confident on these results.

\begin{table}[htbp]
	\caption{The branching fractions of $\Lambda_b^0 \to \Lambda_c^+ \to \ell^- \overline{\nu}_\ell $ in units of $ \% $, where the lower and upper columns in the HBM and LFQM for each lepton pair correspond to the values with and without the hard gluon corrections, respectively.}
	\label{br}
	\begin{tabular}{cccccccccc}
		\hline
		\hline
		Lepton pair&~HBM~&~LFQM~&~Exp~\cite{pdg}~&RQM~\cite{Faustov:2016pal}&LFQM~\cite{Li:2021qod}&LCSR~\cite{Miao:2022bga}&LCSR~\cite{Duan:2022uzm}\\
		\hline
		\multirow{2}{*}{$e^-\overline{\nu}$}&$6.23(58)$&$5.53(77)$&~\multirow{2}{*}{$6.2^{+1.4}_{-1.3}$}~&\multirow{2}{*}{$6.48$}&\multirow{2}{*}{$6.47(96)$}&\multirow{2}{*}{$5.81$}&\multirow{2}{*}{$5.71(98)$}\\
		&$5.69(58)$&$5.35(50)$\\
		\multirow{2}{*}{$\mu^-\overline{\nu}$}&$6.21(57)$&$5.52(77)$&~\multirow{2}{*}{$6.2^{+1.4}_{-1.3}$}~&\multirow{2}{*}{$6.46$}&\multirow{2}{*}{$6.45(95)$}&\multirow{2}{*}{$5.78$}&\multirow{2}{*}{$5.69(98)$}\\
		&$5.67(58)$&$5.33(49)$\\
		\multirow{2}{*}{$\tau^-\overline{\nu}$}&$1.95(11)$&$1.91(23)$&~\multirow{2}{*}{$1.5(4)$}~&\multirow{2}{*}{$2.03$}&\multirow{2}{*}{$1.97(29)$}&\multirow{2}{*}{$1.55$}&\multirow{2}{*}{$1.66(26)$}\\
		&$1.83(12)$&$1.87(15)$\\
		\hline
	\end{tabular}
\end{table} 
	
To see the hard gluon effects, we adopt two different schemes for the form factors.
In the first one, we calculate them directly from the quark models.
In the second one, we plug  the values of  $\overline{\Lambda}$, $b_1$, $b_2$ and $\overline{\xi}(\omega)$ from Table~\ref{IW} into Eq.~\eqref{gluon}.
The branching fractions along with those in the literature are collected  in Table~\ref{br}, where the lower and upper columns in the HBM and LFQM for each lepton pair are the values with and without the hard gluon corrections, respectively, which decrease the branching fractions about $10\%$.
The branching fractions in Table~\ref{br} are consistent with the experimental data and the those in literature, where the uncertainties of our results are smaller than those of other approaches with uncertainties provided.

Beside the integrated branching fractions, the differential decay distributions and other angular observables also provide additional ways to probe the form factors.
They are well discussed in Ref.~\cite{Gutsche:2015mxa}.
For the sake of simplicity, the values of the angular observables have not been given in this work.
We point out that it is possible to fully reconstruct the form factors from the angular observables in the experiments, which is  demonstrated explicitly in Ref.~\cite{BESIII:2022ysa}.

\begin{figure}[h]
		\centering
		\includegraphics[width=0.45\linewidth]{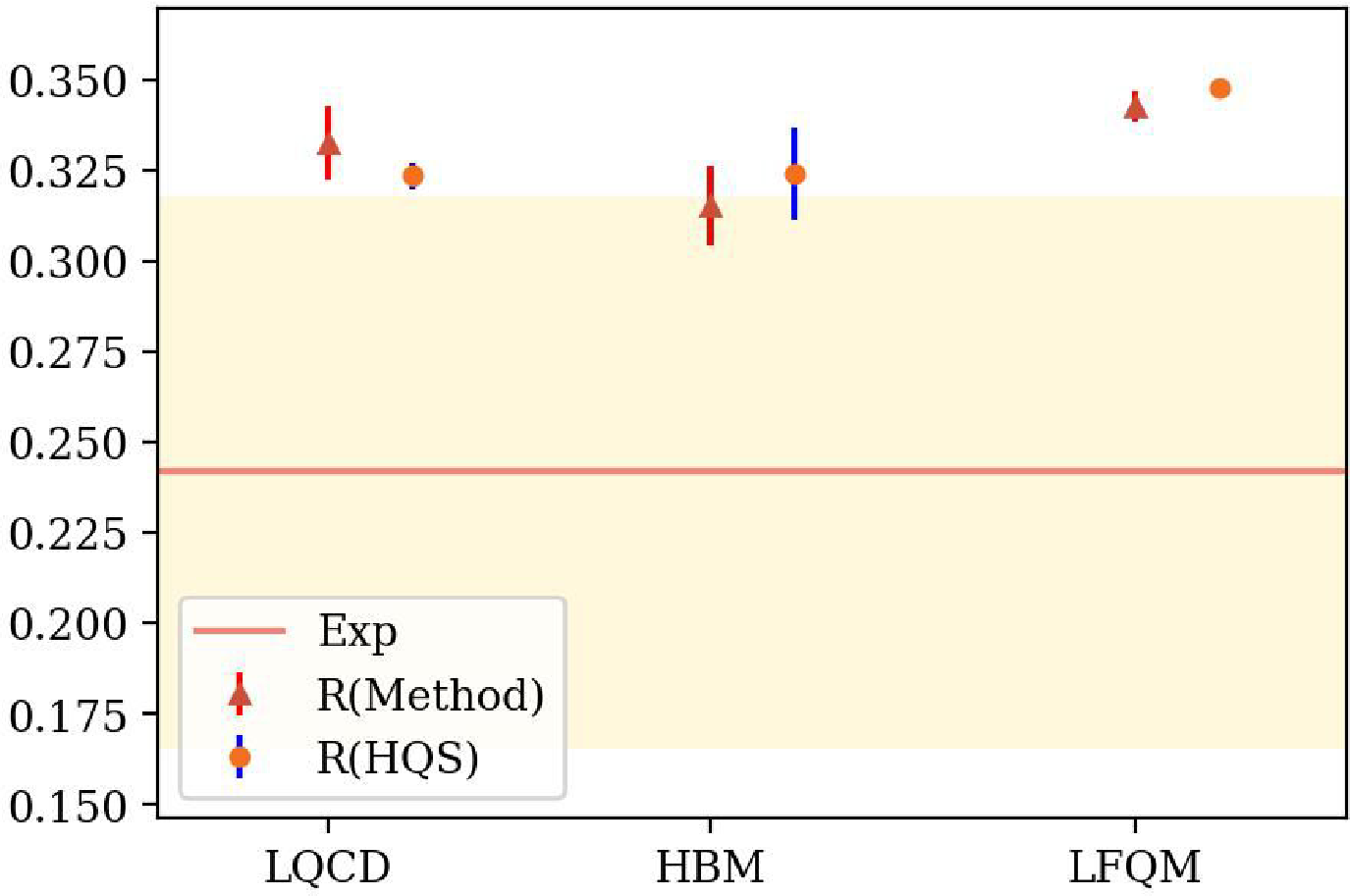}
		\caption{The comparison of $R_{\Lambda_c}$.}
		\label{RLcFig}
	\end{figure}

To further test the lepton universality, we compute the ratio of $R_{\Lambda_c}$, given in Table~\ref{RLctab} and Fig.~\ref{RLcFig}, where $R_{\Lambda_c}(\text{LQCD})$ come from two different works, in which the upper column is purely from the LQCD calculations~\cite{Detmold:2015aaa}, 
while the lower column corresponds to the one which the heavy quark symmetry is imposed to lower the uncertainties~\cite{Bernlochner:2018kxh}. 
The predictions of the HBM fit well with the experimental data, but the ones of the LFQM disagree with the LHCb results in contrast.
	
\begin{table}
	\caption{Comparisons of $R_{\Lambda_c}$ in different approaches.}
	\label{RLctab}
	\begin{tabular}{ccccccc}
		\hline
		\hline
		HBM&LFQM&LQCD~\cite{Bernlochner:2018kxh,Detmold:2015aaa}&~LHCb~\cite{LHCb:2017vhq}&LFQM~\cite{Li:2021qod}&LCSR~\cite{Duan:2022uzm}\\
		\hline
		$0.3154(109)$&$0.3457(70)$&$0.3328(102)$&\multirow{2}{*}{$0.242(76)$}&\multirow{2}{*}{$0.30(9)$}&\multirow{2}{*}{$0.292$}\\
		$0.3243(126)$&$0.3506(46)$&$0.3237(36)$\\
		\hline
		\hline
	\end{tabular}
\end{table}
	
	\section{Conclusion}\label{sec4}
	
We have examined the heavy quark symmetry in the HBM and LFQM.
The inputs of the HBM are fixed from the mass spectra, while the parameters in the LFQM are tightly constrained by the heavy quark symmetry.
We have demonstrated that the two models all respect the heavy quark symmetry.
	
The HQE parameters of $\overline{\Lambda}, b_{1,2}$, $\xi(\omega)$ and $\xi_{ke}(\omega)$ have been computed.
Our results of $\overline{\Lambda}$, $\overline{\xi}(\omega)$ and $b_1$ are compatible with those in the literature, but the sign of $b_2$ is opposite comparing to the LHCb+LQCD results.
Explicitly, we have shown that $(b_1,b_2)=(-0.173,0.518)$ from the LFQM, and $(\overline{b}_1, \overline{b}_2) =(-0.141\pm 0.001, 0.351\pm 0.035)$ from the HBM in units of GeV$^2$. 

We have also calculated ${\cal B}(\Lambda_b^0\to\Lambda_c^+\ell^-\overline{\nu}_\ell)$  and $R_{\Lambda_c}$ with and without the hard gluon corrections.
We have found that the hard gluon corrections  decrease the branching fractions about $10\%$.
Explicitly, we have obtained that ${\cal B}(\Lambda_b^0\to\Lambda_c^+e^-\overline{\nu}_e)=(5.69\pm 0.58, 5.35\pm 0.50)$, ${\cal B}(\Lambda_b^0\to\Lambda_c^+\mu^-\overline{\nu}_\mu)=(5.67\pm 0.58 $, $5.33\pm 0.49)$, and $R_{\Lambda_c}=(0.3243\pm 0.0126$, $ 0.3506\pm 0.0046)$, for the results of (HBM, LFQM), respectively.
Our predicted values of the branching fractions show good consistencies with the experimental data of results ${\cal B}(\Lambda_b^0\to \Lambda_c^+ e^-\overline{\nu}_e, \tau^- \overline{\nu}_\tau) = (6.2^{+1.4}_{-1.3}\,, 1.5\pm 0.4)\%$.

\section*{Acknowledgments}
	This work is supported in part by the National Key Research and Development Program of China under Grant No. 2020YFC2201501 and  the National Natural Science Foundation of China (NSFC) under Grant No. 12147103.

\end{document}